\documentstyle[bo99,epsfig]{article}

\title{BeppoSAX Observations of Bright Radio Galaxies}

\author{Paola Grandi}

\affil{Istituto di Astrofisica Spaziale - CNR, Via Fosso del 
Cavaliere 100, I-00133, Roma, Italy}                                                
\begin{document}

\maketitle

\begin{abstract}
BeppoSAX observations of Broad Line Radio Galaxies (BLRGs) have shown that 
they have a considerable variety of spectral properties
and important differences with respect to their radio-quiet counter-part
Seyfert 1s.
In radio galaxies the soft photons are often absorbed by cold material.
In contrast, in Seyfert 1s the soft photons are generally
absorbed by warm gas. 
The iron lines, always detected in Seyfert 1s, are not always present 
in BLRGs and generally are weak. In addition, small iron line
equivalent widths seem to correspond to weak reflection components in 
radio galaxies.

The emerging picture of BLRGs  is complex. 
Probably several X-ray components, 
jet, accretion flow and molecular torus, mixed in different way 
in different objects,  contribute to the 
production of their X-ray spectrum and determine the observed variety.
The weakness of the reprocessed features can 
be explained either by a dilution of the Seyfert-like continuum from 
non-thermal (jet) radiation or by an accretion gas that is 
hot and geometrically thick close to the black hole and cold geometrically 
thin (i.e. able to reprocess the primary X-ray radiation) at larger radii.
\end{abstract}

\section{Introduction}

The most basic classification of Active Galactic Nuclei (AGN) 
consists in dividing them in two classes: radio-quiet and radio-loud. 
This represents not only 
an observational distinction but a basic physical difference 
whose basis is still not understood. Radio morphologies 
(lobes, jets etc.) are obviously the 
product of some physical mechanisms at work in the nuclei of radio-loud 
objects. 
Rees et al. (1982) suggested, for example, that the primary engine in radio 
galaxies was a spinning black hole fed by thick, and hot, accretion flow.
Later, Blandford (1990) and Meier (1999) have explicitly identified 
the black hole spin as a possible physical parameter responsible for the 
radio-loud and radio-quiet dichotomy. If the hole rotates
faster, it is more efficient in producing the jets observed in 
radio-loud objects.

X-ray photons seem to be the best probe to investigate the radio loudness 
issue as they are produced (and reprocessed) 
in the inner regions of AGNs where accretion occurs. 
X-ray observations of radio-loud objects therefore provide unique 
data to test  the model hypotheses. 
In particular, two questions can be addressed:
1) Does the jet contribute to the X-ray emission in non-blazar radio-loud 
AGNs? 2) Are the same accretion processes at work in radio galaxies 
(and in general in radio-loud AGNs) and Seyfert galaxies?

BeppoSAX is a broad band (0.1-200 keV) X-ray satellite (Scarsi 1993) 
which has ideal characteristics for this purpose.
Here we present a BeppoSAX analysis of 6 Broad Line Radio Galaxies 
and compare our sources with a sample of 13 Seyfert 1 galaxies
also observed by BeppoSAX (Matt these proceedings, hereafter M00).
Based on the properties of optical-UV spectra, 
BLRGs are considered the radio-loud counterpart of Seyfert 1s  
(Urry and Padovani 1995).
The comparison of the X-ray (nuclear) properties of these two classes 
of objects is then 
particularly appropriate to investigate the radio-loud and radio-quiet 
dichotomy.

\section{Observations and Results}
 
The radio galaxies presented  here are part of a larger sample of 
Radio-Loud Emission-Line AGNs observed by BeppoSAX, whose analysis is still
on-going.  In this paper, we restrict the discussion to 
a subsample of 6 BLRGs, which can be directly compared with 
the sample of Seyfert 1s discussed by M00. The sample is shown in Table 1.
All the 3C sources have been observed for about 100 ksec. 
PKS2152-69 and Pictor A were observed for $\sim17$ and $30$ ksec, respectively.
\begin{table}
\begin{center}
\footnotesize
\caption{BeppoSAX Sample of Broad Line Radio Galaxies}
\begin{tabular}{lcccc} 
\hline 
&&&&\\
Source & z & N$_{Gal}$& Radio$^a$  & L$^b_{2-10 keV}$  \\
       &   & 10$^{21}$ cm$^{-2}$& Morphology & $\times10^{44}$ \\
&&&&\\
PKS2152-69       & 0.027 & 0.25 & FRI/FRII   & 0.2 \\
Pictor A         & 0.035  & 0.42& FRII      & 1.0  \\
3C120            & 0.033  & 1.1 &FRI/S       & 2.4   \\
3C111            &0.048  & 3.2 &FRII/S     & 2.9  \\
3C390.3          &0.057  & 0.42 &FRII/S     & 3.3   \\
3C382            & 0.059 & 0.88& FRII      & 9.4 \\
\hline\\
\multicolumn{5}{l}{$^a$ -- FR = Fanaroff-Riley, S = superluminal source}\\
\multicolumn{5}{l}{ $^b$ -- Luminosity corrected for 
absorption in ergs sec$^{-1}$}\\
\end{tabular}
\end{center}
\end{table}

\subsection{BLRG Spectral Analysis}
The spectral analysis results, obtained by 
simultaneously fitting the LECS (0.1-4 keV) MECS (1.5-10 keV) 
and PDS (15-100 keV) instruments are reported in Table 2.

A simple power law plus Galactic absorption gave 
good fits to the data of PKS2152-69 and Pictor A.
The iron line is absent in Pictor A in agreement with the ASCA measurement
(Eracleous et al. 1998).
For PKS2152-69, the exposure time was too short to allow the detection of even a strong feature.
For both, we could obtain only an upper limit for the iron
line equivalent width. 
In 3C111 the continuum was also  a
simple power law and the iron line only marginally detected (in spite of 
the long exposure time).  
The soft photons appeared  strongly absorbed. It is however possible that the 
cold absorber is not intrinsic to the source but associated to our Galaxy.
3C111  is in fact behind the  Galactic dark cloud, Taurus B 
(see Reynolds et al. 1998 for more details).

More complex models were necessary to fit the other radio galaxies.
Absorption in excess of the Galactic column density, iron lines, reflection (Ref.) 
humps, soft excesses and bendings of the high energy spectrum (Cutoff)
are common spectral features, although not always simultaneously present in 
each source. 
\begin{table}
\footnotesize
\caption{BLRG SPECTRAL FIT RESULTS}
\begin{tabular}{lccccccc} 
\hline
&&&&&&&\\
Source & $\Gamma$ & N$_H^b$ & Refl. & Cutoff
& E$_{Fe}$ & $\sigma$ &  EW \\
          &          &  & &  keV&
keV & keV & eV \\
&&&&&&&\\
PKS2152-69$^{a}$ &  1.79$\pm0.1$& 0.25 (f) &... &... &
6.4 (f) & 0.0 (f) & $<429$ \\
Pictor A & 1.63$\pm0.06$& 0.42 (f)&...  & ...  &
6.4 (f) & 0 (f) & $<102$\\
3C120$^{\star}$ & 1.80$^{+0.12}_{-0.30}$ & 2.4$^{+1.8}_{-1.3}$ & 0.7$\pm0
.4$ & 109$^{+125}_{-77}$ & 
6.2$\pm0.4$ & 0.27$^{+0.40}_{-0.27}$ &59$^{+82}_{-42}$\\
3C111$^{\dag}$& 1.65$\pm0.04$ &7.1$^{+0.9}_{-0.8}$ &$<0.3$ & $>90$ &
6.6$^{+0.4}_{-0.2}$ & 0 (fixed) & 58$^{+31}_{-55}$\\
3C390.3& 1.80$^{+0.05}_{-0.04}$ & 1.3$\pm0.2$ & 1.2$^{+0.4}_{-0.3}$& $>12
3$&6.4$\pm0.1$ & 0.07$^{+0.21}_{-0.07}$& 136$^{+40}_{-36}$\\
3C382$^{\star}$& 1.79$\pm0.04$& 0.88 (f) & 0.4$^{+0.3}_{-0.2}$&
155$^{+148}_{-59}$&
6.5$^{+0.9}_{-0.2}$& 0.00 (f) &31$^{+44}_{-15}$  \\
\hline\\
&&&&&&&\\
\multicolumn{8}{l}{$^a$ -- MECS data only; $^b$ -- in unit 
of $10^{21}$ cm$^{-2}$}\\
\multicolumn{8}{l}{$^\star$ -- Sources with detected soft excess; 
$^\dag$ -- Iron line marginal detection}\\
\end{tabular}
\end{table}
\normalsize

The soft excess was detected in two sources, namely 3C120 and 3C382.
It was  parameterized with a steep power law ($\Gamma^{\it soft}=3-4$)
and contributed to the total emission in the 0.1-2 keV band
more than 60$\%$ in both the sources.
The origin of this excess is not clear. It could be related to 
thermal emission from a cold thin disk, to radiation coming 
by extended thin plasma as well as to a very soft jet emission.

All the BLRGs with detected iron line show reflection components.
Although the strength of the reflection (Ref.) is not very well constrained, 
data seem to suggest that radio-galaxies with weak iron lines have also 
weak reflections (Figure 1). 
If confirmed, this would imply that 
the line is generated in the same material which reflects the X-ray 
primary photons.  

\begin{figure}
\centerline{\psfig{figure=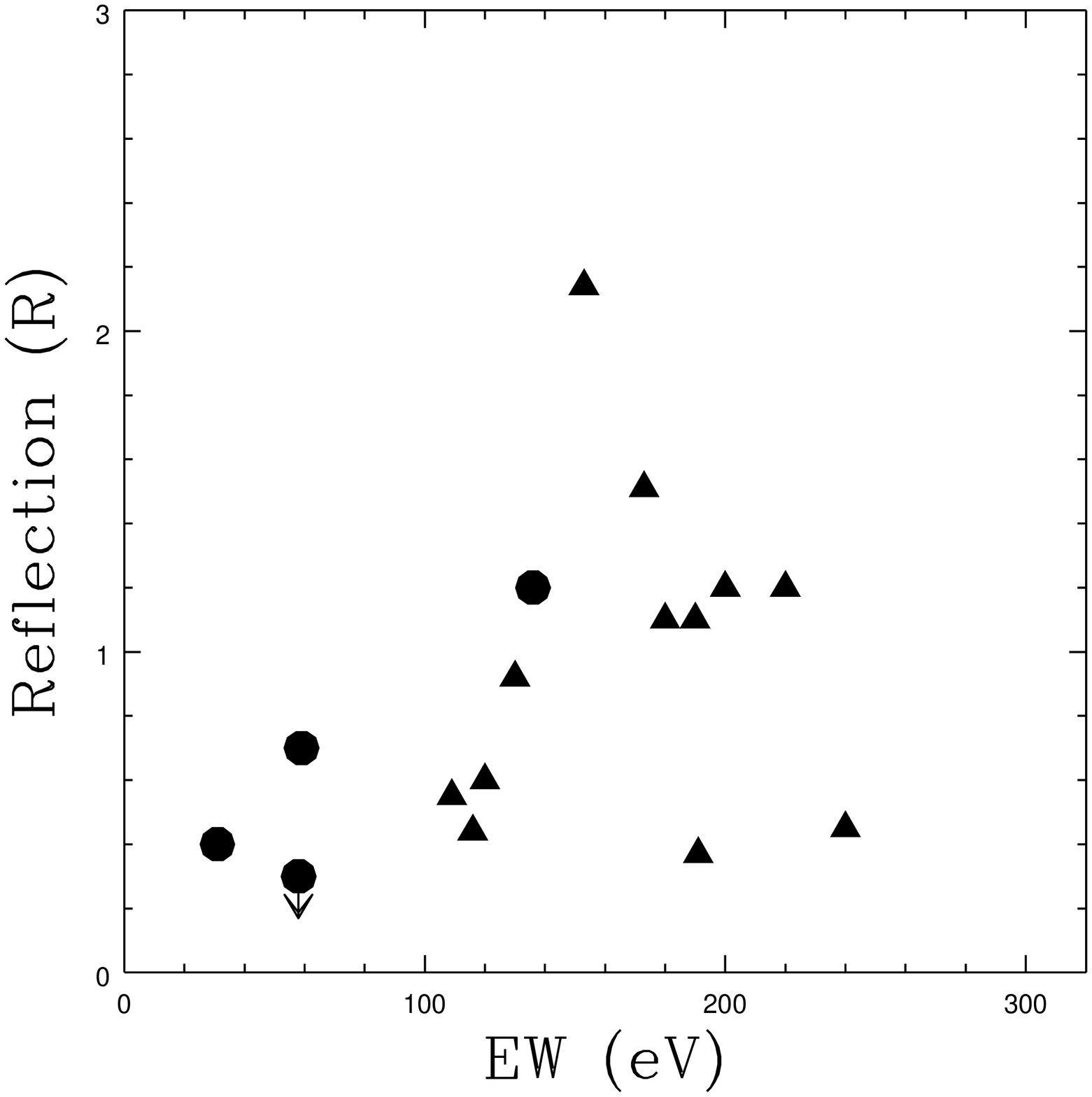,height=2.5in,width=2.5in}
\psfig{figure=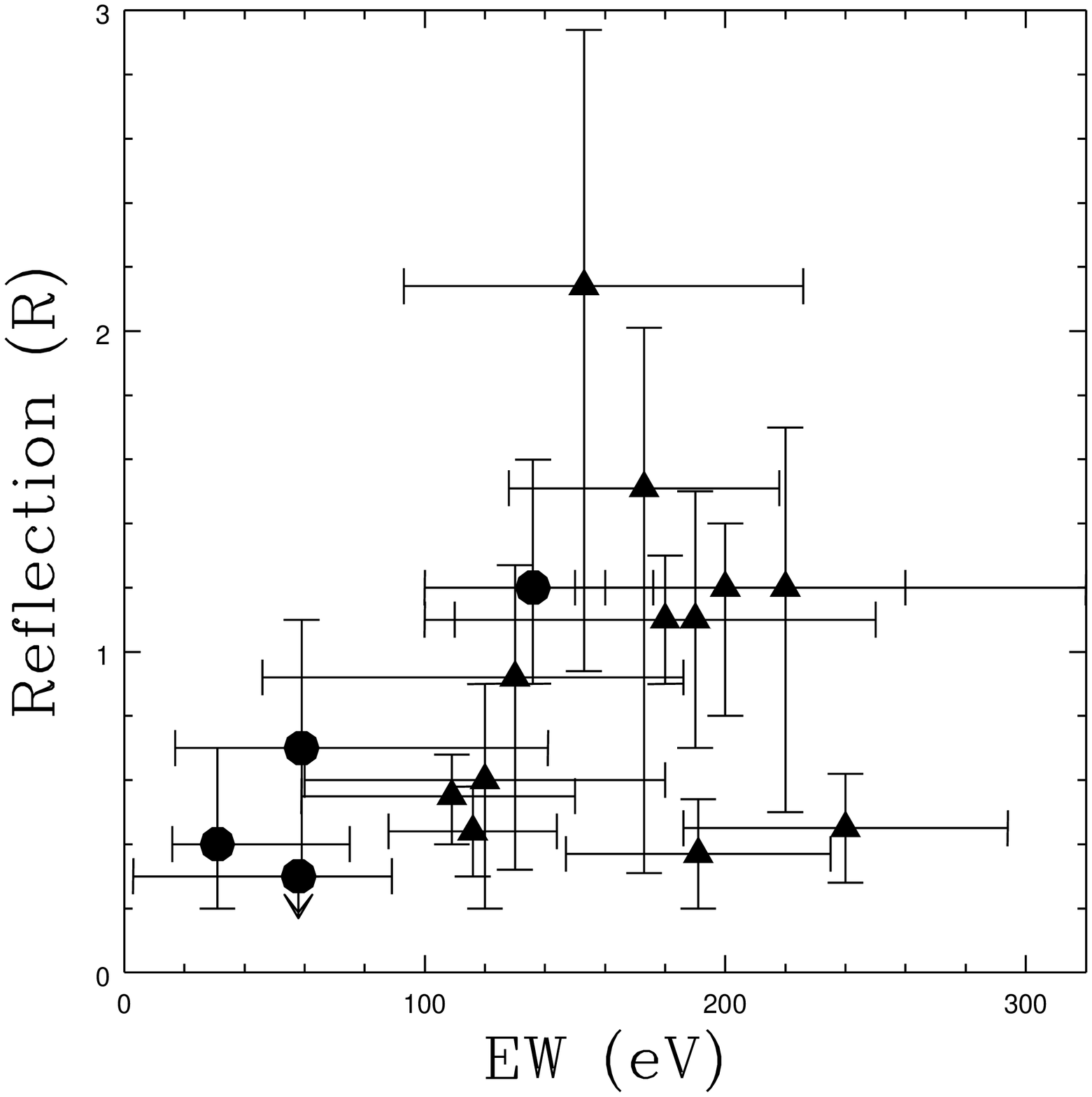,height=2.5in,width=2.5in}}
\caption{({\it left panel}) -- The amount of 
reflected radiation (Reflection) is 
plotted as a function of the strength of the iron line (EW)
for the BLRG sample (circles). For comparison the Seyfert 1 (triangles)
sample of M00 is also shown. Note that the radio-loud AGN are characterized by 
generally weaker iron lines and reflection components. 
({\it right panel}) -- Same plot with error bars.}
\end{figure}

\subsection{Comparison between BLRGs and Seyfert 1s}

We compared the spectral properties of our BLRGs with a sample of 13 
Seyfert 1s observed by BeppoSAX (M00).
The 2-10 keV luminosities of Seyferts 
(ranging from $\sim10^{42}$ to  $10^{44}$ erg sec$^{-1}$) and BLRGs (see Table 1) partially overlap. 

The X-ray primary power law of BLRGs is generally flat ($<\Gamma^{\it BLRG}= 
1.73>$, rms dispersion $\sigma^{\it BLRG}_{rms}= 0.08$).
The Seyfert 1 sample is characterized by 
a steeper  average spectral slope.
However, the larger spread of the spectral indices 
in radio-quiet AGNs  ($<\Gamma^{\it Sey~1}>$=1.87, 
$\sigma^{\it Sey~1}_{rms}=0.24$), does not allow to statistically 
confirm any difference between Seyfert 1s and BLRGs.

Half the 3C sources in our sample shows a bending of the X-ray spectrum at 
high energies. When it is modeled with an exponential cutoff, the 
cutoff energies are similar to those observed in Seyferts. 

While in radio-quiets the iron lines and the reflection components 
are always present, in BLRGs the reprocessed features are 
detected in only 3 sources. 
It should be noted the absence of the lines cannot be only attributed to poor statistics.
In addition in radio-louds, the iron line equivalent widths
are significantly smaller than in radio-quiets 
($\langle EW^{\it BLRG}\rangle =71$~eV, $\sigma^{\it BLRG}_{rms}=45$~eV;
$\langle EW^{\it Sey1}\rangle =174$~eV, 
$\sigma^{\it Sey1}_{rms}=57$~eV).

A general weakness of the reprocessed features in BLRGs 
is also confirmed by the XTE results presented by Sambruna and Eracleous 
(these proceedings).
Note that Wozniak et al. (1998) have already discussed this possibility 
analyzing ASCA, GINGA and OSSE data.

A cold absorbing column in excess of Galactic is observed in 3 
(including 3C111) out of 5 BLRGs with LECS data.
No source shows absorption edges typical of warm absorber, 
which, on the contrary, is rather common in Seyfert 1s. 
It is possible that the absorbing material is different in BLRGs and 
Seyfert 1s, being warm in radio-quiets and cold in radio-louds
(see also Sambruna Eracleous and Mushotzky 1999).
This is also supported by a historical study of the column density 
changes in 3C390.3.
The long-time (years) variability of the 
intrinsic N$_H$ does not appear to be correlated to the flux intensity at 
1~keV (see Fig.~2 in Grandi et al. 1999).
It is possible that variations in the geometry 
of the absorber rather than changes of its ionization state 
(as expected in the case of a warm absorber)
are responsible for the N$_H$ long term variability.

\section{Discussion }

Two important points arise from the BeppoSAX study of BLRGs:
1) there is not a unique type of BLRG X-ray spectrum, but a 
variety of cases; 2) there are several differences between
BLRGs and Seyfert 1 galaxies. The most impressive difference concerns
the reprocessed features that are weaker in radio-louds 
than in radio-quiets (Fig. 1). How can these results be explained?

\noindent
BLRGs are complex objects, in which at least three X-ray 
components, a jet, an accretion flow
and a molecular torus, can contribute to the formation of the spectrum.

\noindent
{\bf Jet} --
In some BLRGs the radio jet shows superluminal motion. It is then reasonable
to suppose that Doppler-enhanced radiation
also contaminates the X-ray spectra and dilutes the reprocessed 
spectral features when the Doppler factor of the jet 
($\delta$=[$\gamma$(1-$\beta$ cos)]$^{-1}$) is sufficiently large.
If, in agreement with the AGN Unified Schemes, BLRGs
and intrinsically powerful blazars 
are the same objects seen at different angles of view, 
also the jet output of radio galaxies should be Inverse Compton 
in the PDS band (Fossati et al. 1998, Ghisellini 1998).
If this is the case, it is rather improbable that the observed 
high energy  steepening in BLRGs is related to the jet.
In powerful blazars, the Compton break (i.e. the Compton peak in the Spectral 
Energy Distribution, SED) is usually revealed at MeV-GeV energies.
In order to occur at 100-200 keV as observed in BLRGs, the Doppler factor 
( $\delta$=[$\gamma$(1-$\beta$ cos)]$^{-1}$) 
should be smaller by about a factor 100 or more.
This would imply a strong de-amplification of the non-thermal radiation 
($I^{obs}$($\nu$)=$\delta^3$I$^{intr.}$($\nu')$) and, in turn, 
a difficult detection of the jet.

\noindent
{\bf Accretion  Flow} --It is not really clear how gas accretion 
occurs in radio-loud objects.
The simplest  approach is to extend the physical models developed 
for Seyferts to Broad Line Radio Galaxies.
As in the case of radio-quiet AGNs, the accretion gas flow could be
a cold geometrically thin disk with a hot corona above it 
(Haardt and Maraschi 1991, 1993; Petrucci et al. 2000).
The role of the corona is to transform into X-ray photons, via 
inverse Compton scattering, the UV photons generated by the disk.
Down-scattered X-rays, in turn, hit the disk and are
reprocessed, generating an iron line and a reflection hump above 10~keV.
A  Seyfert-like accretion disk should then produce the following signatures
in X-ray BLRG spectra: a soft excess (related to the thermal disk emission), 
a high energy thermal cutoff (related to the corona temperature), a 
broad red-shifted iron line (which suffers the vicinity of the 
strong black hole gravitational field) and a strong reflection component 
(produced by a disk subtending a $\sim 2\pi$ solid angle to the X-ray primary 
source, i.e. Refl$\sim \Omega/2\pi\sim1$). 

Alternatively, one can assume
that different physical/geometrical accretion configurations 
occur in AGNs which discharge large amount of energy
in jets. 
Rees et al. (1982) speculated on the possibility that in radio-galaxies 
the hot accreting gas is in the shape of an ion-supported torus characterized by low radiative efficiency.
The Advection Dominated Accretion Flow (ADAF) models
(Narayan e al. 1998), recently proposed, follow similar lines of though.
Shapiro Lightman and Eardley (SLE, 1976) found another solution for the 
hot flow, that resembles the ion-supported torus.
However,  in the SLE model the energy produced in the accreting gas
by viscosity is locally radiated and not advected radially.

Then the accretion flow in BLRGs
could be  hot and geometrically thick in the 
inner region and become cold and geometrically thin only at larger radii 
(Chen and Halpern 1989). 
\begin{figure}[t]
\centerline{\epsfig{file=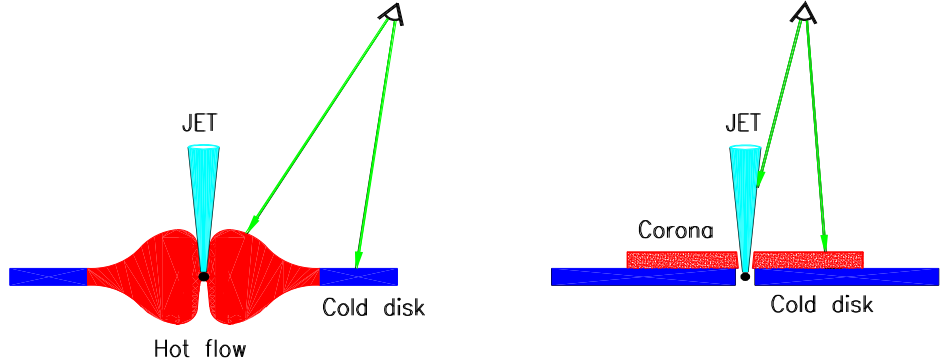, height=2.5in, width=5.0in}}
\caption[]{A schematic diagram of possible radio-loud accretion flows.
The accretion can be hot geometrically thick near the black hole
and cold geometrically thin in the outer regions (left panel). The weakness of the reprocessed features is assured by the small solid angle subtended by cold
material. In this model, the jet radiation can be negligible. 
If accreting processes are the same in radio-louds and radio-quiets AGNs, the accretion flow is a unique cold and optically thin disk (right panel). 
In this case, a jet
contribute is necessary to explain the weak iron lines and reflections components observed in BLRGs.}
\end{figure}
Given the smaller covering factor of the cold (and reprocessing) matter to the
X-ray primary source, 
this accretion configuration predicts
less prominent soft excesses, narrower weaker 
iron lines and smaller reflection components than the Seyfert case.
In addition, a lack of correlation between the variations of the primary X-ray 
source and the reprocessed features is expected; the entity of the temporal 
delay depending on the inner radius of the cold disk. 

\noindent
{\bf Molecular Torus} -- It is probable that a  thick wall of 
absorbing material (perhaps in a toroidal shape) surrounds the accretion flow.
The torus can produce the iron line and reflect the X-ray primary 
photons (Ghisellini Haardt and Matt 1994) further complicating
the X-ray spectrum.  In addition, if the opening angle of the torus is 
small, its (less thick) upper 
layers could be intercepted by the observer  
and cause the soft X-ray depletion sometimes observed in BLRGs.
 
Since its distance from the X-ray source is large ($\geq 1$
pc), iron line and reflection components should respond with a considerable 
delay (light-years) to the continuum variations.

\vspace{0.3cm}
\noindent
It is possible and also probable that all these components are present 
in BLRGs. If they are mixed in different ways in different objects, 
the variety of X-ray spectra observed with BeppoSAX would be easily explained.
As shown in Table 3  the main BeppoSAX spectral features 
of the BLRGS with high signal-to noise spectra
(i.e. pointed for about 100 ksec) can be opportunely 
reproduced choosing a plausible combination of the nuclear components.

\begin{table}
\begin{center}
\footnotesize
\caption{BLRG NUCLEAR COMPONENTS}
\begin{tabular}{ccc} 
\hline
&&\\
Source & BeppoSAX Spectral characteristics & Dominant Spectral Components\\
&&\\
3C120 & soft excess                  & Jet + Seyfert-like disk\\
      & weak (broad?) Fe line & \\
      & reflection&\\
      & high energy cutoff & \\   
&&\\
3C111 & iron line (?)          & Jet \\
      & no reflection          & \\
      & no cutoff              & \\ 
&&\\
3C390.3 & narrow strong Fe line &  Hot flow + cold disk + Torus\\ 
        & strong reflection       &\\
        & no soft excess          & \\
&&\\
3C382   & narrow weak Fe line     & Hot flow + Torus  \\
        & weak reflection         & (+ external cold disk?) \\
        & energy cutoff           & \\
        & soft excess &\\  \hline
\end{tabular}
\end{center}
\end{table}
\normalsize

This tentative table shows that a Seyfert-like disk alone can not 
reproduce the observations. The weakness of the reprocessed features 
requires the presence either of a non-thermal radiation from a jet or 
of a cold disk sub-tending a small solid angle to the X-ray primary source
(see figure 2).

\section{Seyfert-like or ADAF-like Accretion?}

The sources in Table 3 which seem better to fit
the Seyfert-like+jet model are 3C111 and 3C120.
If the model is correct,  the weak (but detected) iron line in 
3C120 assures that Seyfert and Blazar components  have comparable intensity 
around $6$ keV, i.e. the non-thermal power law is important and
can effect the continuum shape.
We then tested whether the 3C120 spectrum 
can be fitted by a mix of beamed and un-beamed 
radiation on the entire BeppoSAX band (i.e. on about 3 decades in energies).

A simple power law was assumed to represent the jet (blazar) spectrum and a 
power law with cutoff plus iron line and reflection component were 
utilized to mimic the Seyfert emission. The Seyfert equivalent width 
was assumed EW$^{sey}=174$ eV, the average value from the Matt sample.
Different combinations of spectral slopes for the Seyfert and blazar 
models were tested. In all the cases, 
the relative normalization at 6.4 keV of the two power laws
was fixed in order to reproduce the observed Fe equivalent width 
reported in Table 2.
As shown in Figure 3 ({\it left panel}), 3C120 data can be fitted by 
a blazar and a Seyfert-like power law (in Figure $\Gamma^{Blazar}=2$, 
$\Gamma^{Sey}=1.7$).
However the model requires very low energy cutoff 
($E_{Cut}\sim 70$ keV) and a strong  reflection (Ref$\sim 2$).
These  best fit values are rather extreme if compared to those of Seyferts.
This slightly disfavors the jet model, although  it cannot be 
completely rejected given the large uncertainties associated to 
the parameters.

A geometry that assumes a hot inner flow and cold disk surrounding it
seems to be particularly appropriate for 3C390.3. 
In this source the UV bump is weak,
there is not soft excess and the iron line is narrow.
In this case a  molecular torus (or a warped disk) also contributes to 
increase the strength of the iron line and the reflection 
component (Grandi et al. 1999).

The idea that the iron line emitting regions cannot be located near the X-ray 
is also supported by recent BeppoSAX observations
of the very bright radio-galaxy Centaurus A (Cen A)
($F_{\it~2-10~keV}\sim 10^{-10}$ erg  cm$^{-2}$ sec$^{-1}$). 
It is a FR I optically classified as Narrow Line Radio Galaxy and shows 
a radio-optically-X-ray jet pointed far away from 
us ($i\sim 60-70^{0}$).
It is characterized by a X-ray spectrum rather complex. However above 3 keV
it is dominated by the nuclear point-like source (Turner et al. 1998).

Cen A was observed on 1997 February 20-21 (30 ksec), 
on 1998 January 6-7 (50 ksec) and twice for 40 ksec during the 1999 summer
on July 10-11 and August 2-3.
The repeated observations of this radio-galaxy, thanks to 
their high statistics, 
have allowed a detailed study 
of the nuclear flux variations versus the iron line flux changes.
As clearly shown in Figure 3 ({\it right panel}), line and continuum do not
vary together, in particular the iron line was more intense 
when the source was weaker (compare the 1997 and 1998 observations).
Since the inclination of the jet is large in Cen A, the contribution 
of non-thermal radiation to the total X-ray nuclear continuum is expected to 
be negligible.
Then Figure 3 ({\it right panel}) 
simply indicates that the  line emitting region 
responds with a significant delay to the continuum variations, 
as it is expected if the reprocessing gas is located far from the primary 
X-ray source. 
\begin{figure}
\centerline{\psfig{figure=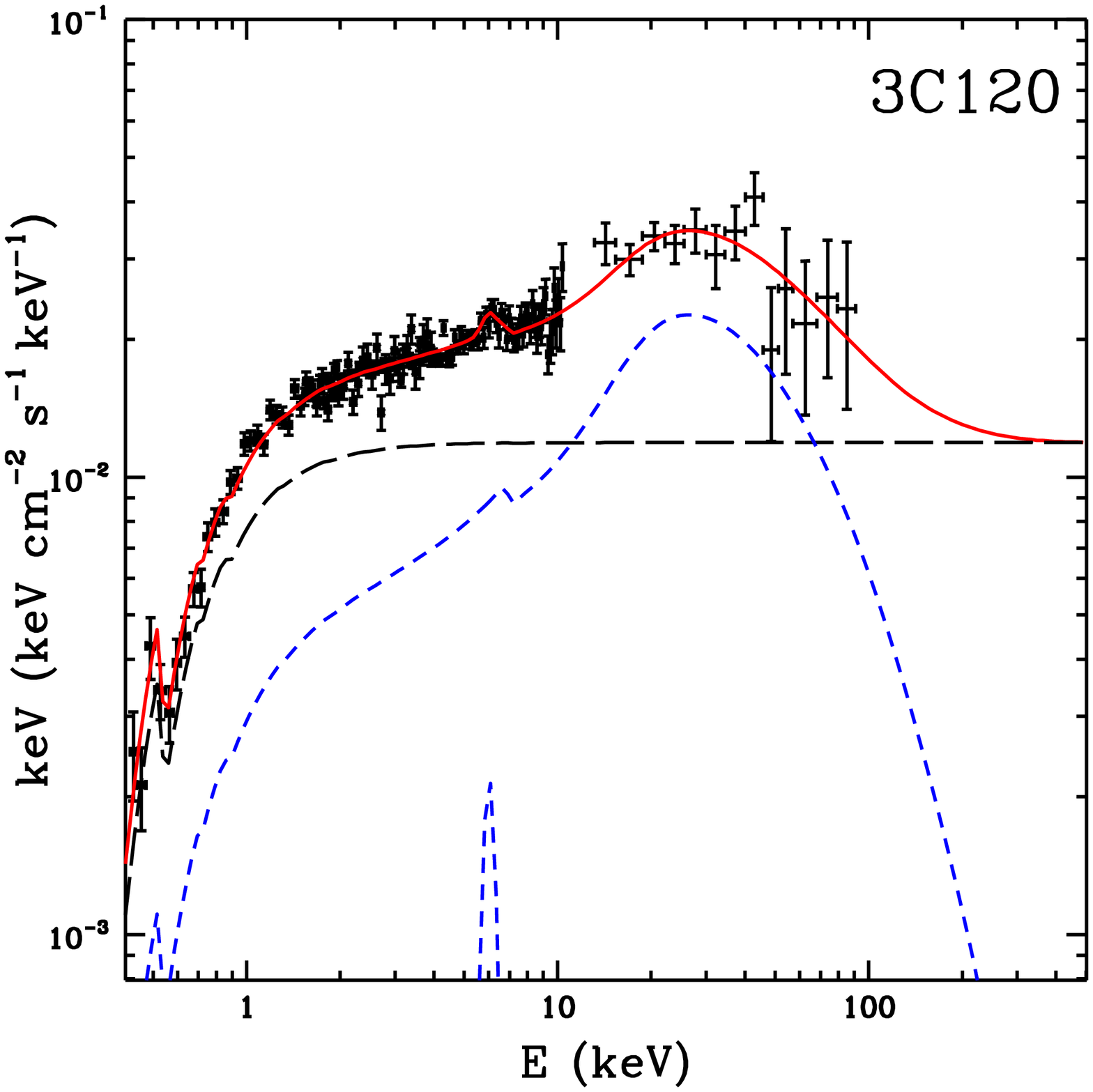,height=2.0in,width=2.5in}
\psfig{figure=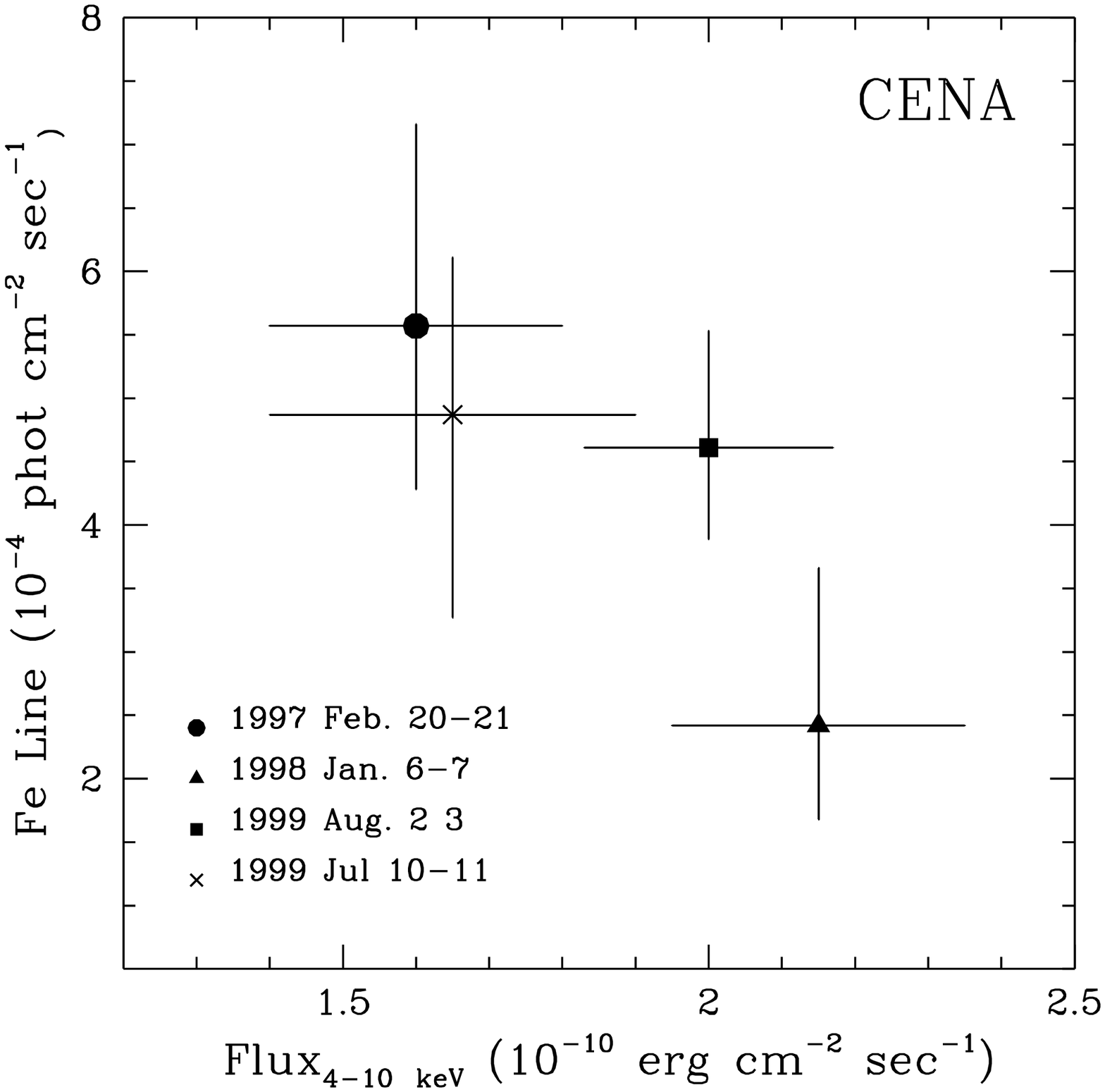,height=2.0in,width=2.5in}}
\caption{({\it left panel}) -- 3C120: the 3C120 data can be fitted with a mix 
of Seyfert-like (red line) and blazar-like (green line) spectra.
({\it right panel}) -- CenA: the variations of the Fe line intensity 
are reported as a function of the 4-10 keV nuclear flux changes. The iron line and the continuum do not change together.}
\end{figure}

We conclude that, although, in principle, the two proposed accretion 
scenarios are both viable, the idea that a jet reduces 
the reprocessed features is lightly disfavored.

\section{Conclusions}

BeppoSAX analysis of 6 BLRGs has shown that  
a variety of X-ray spectra exits. A  mix of 
different components, a jet, an accretion flow and a molecular torus,  
could explain the observations.
The comparison between our 6 BLRG and a sample of 13 Seyfert 1s 
has pointed out important differences between radio-loud and radio-quiet AGNs.
In particular, it has been shown that  BLRG reprocessed features
are often absent or weaker than in Seyfert 1s.
This result has important implications.
If the accretion flow in BLRGs and Seyferts is identical 
(i.e a cold thin optically thin disk with a hot above it)
a strong contribution by Doppler-enhanced radiation 
is necessary to explain the weakness of the reprocessed features.
Alternatively, if the jet emission is not important, the cold re-processing
gas has to subtend a smaller solid angle to the X-ray primary source.
A possibility is that the X-ray continuum is produced by 
a hot geometrically thick ion-supported torus that illuminates a cold 
thin disk at larger radii. 

Choosing between the two scenarios is still premature.
The new satellites (XMM, Chandra, INTEGRAL) will play an important 
role in solving the problem.
Detailed studies of the iron line profiles in a large (and well selected)
sample will allow to confirm the presence of cold matter very near 
to the black hole (if lines are broad and  redshifted) or at large distances 
(if lines are narrow).

Variability studies will be also crucial.
An anti-correlation is expected between the EW and the X-ray continuum 
flux, if the jet is dominant: the observed continuum (disk + jet) should be
more variable than the Fe line (from disk).
Alternatively, if no correlation is found, the Fe variations do not follow
the continuum flux variations and 
a temporal delay effect (as observed in CenA case) is present.
Finally,  detailed studies of the hard X-ray continuum will be able to 
detect any beamed radiation.
At very high energies, where the Seyfert-like 
power law drops, the jet, if intense,  should emerge and become 
directly detectable (see Figure 3 {\it left panel}).

\begin{acknowledgements}
I would like to thank all the people who, working with me on 
the BeppoSAX radio-galaxy project, have allowed the writing of
this paper: L. Maraschi, C. M. Urry, E. Massaro,
G. Matt, M. Guainazzi, F. Haardt, P. Giommi, G.G. Palumbo, G. Malaguti. 
An acknowledgment also go the L. Piro and the BeppoSAX CDC team for their 
support to this project.
I also thank G. Ghisellini, L. Ferretti and C. G. Perola 
for the useful comments and stimulating discussions and A. Bazzano and
A.J. Bird for critical reading of the manuscript.
I am very grateful to M. Frutti for invaluable help in realizing Figure 2.
 
\end{acknowledgements}

\end{document}